\documentclass[prl,superscriptaddress, twocolumn]{revtex4}
\usepackage{color}
\usepackage{amsfonts,amssymb,amsmath}
\usepackage[]{graphics,graphicx,epsfig}
\usepackage{amsthm, subfigure}
\usepackage{natbib}

\input{Qcircuit}

\newcommand{\tr}[0]{\mathrm{tr}}

\begin{document}

\title{Quantum computing with black-box subroutines}

\author{Jayne Thompson}
\affiliation{Centre for Quantum Technologies,
National University of Singapore, 3 Science Drive 2, 117543 Singapore,
Singapore}

\author{Mile Gu}
\email{ceptryn@gmail.com}
\affiliation{Center for Quantum Information, Institute for Interdisciplinary  Information Sciences, Tsinghua University, Beijing, 100084, China}
\affiliation{Centre for Quantum Technologies, National University of Singapore, 3 Science Drive 2, 117543 Singapore, Singapore}

\author{Kavan Modi}
\affiliation{Department of Physics, University of Oxford, Clarendon Laboratory, Oxford, OX1 3PU, United Kingdom}
\affiliation{Centre for Quantum Technologies, National University of Singapore, 3 Science Drive 2, 117543 Singapore, Singapore}

\author{Vlatko Vedral}
\affiliation{Department of Physics, University of Oxford, Clarendon Laboratory, Oxford, OX1 3PU, United Kingdom}
\affiliation{Centre for Quantum Technologies, National University of Singapore, 3 Science Drive 2, 117543 Singapore, Singapore}
\affiliation{Department of Physics, National University of Singapore, 2 Science Drive 3, 117551 Singapore, Singapore}

\begin{abstract}
Modern programming relies on our ability to treat preprogrammed functions as black boxes - we can invoke them as subroutines without knowing their physical implementation. Here we show it is generally impossible to execute an unknown quantum subroutine. This, as a special case, forbids applying black-box subroutines conditioned on an ancillary qubit. We explore how this limits many quantum algorithms - forcing their circuit implementation to be individually tailored to specific inputs and inducing failure if these inputs are not known in advance. We present a method to avoid this situation for certain computational problems. We apply this method to enhance existing quantum factoring algorithms; reducing their complexity, and the extent to which they need to be tailored to factor specific numbers. Thus, we highlight a natural property of classical information that fails in the advent of quantum logic; and simultaneously demonstrate how to mitigate its effects in practical situations.
\end{abstract}
\maketitle

The solution of complicated computational problems is typically split into a sequence of subroutines that solve smaller problems. One can construct the desired solution without the necessity of understanding the detailed physical implementation of each individual component. This modularity is of particular importance in quantum computing, where quantum speedup often exploits our ability to encode unitary matrices $U$ of exponential size--within polynomial sized quantum circuits~\cite{Knill,Loyd_Linear}. The execution of such circuits as subroutines allows us to efficiently compute some otherwise intractable functions of $U$.

Deterministic quantum computing with one qubit (DQC1), which efficiently evaluates the trace of an exponentially large unitary matrix $U$, presents an archetypical example~\cite{Knill, DQC1}. Whereas a classical algorithm is forced to access an exponentially large string of numbers, and thus requires exponential time, certain $U$ can be represented with a polynomial sequence of elementary (one or two-qubit) quantum gates. In probing the properties of this circuit by applying it on a maximally mixed register conditioned on an ancillary qubit, one can determine the trace of $U$ efficiently. Indeed, this strategy underpins several important quantum protocols~\cite{dqc1-1, dqc1-2, dqc1-3, dqc1-4}, all of which estimate properties of a given quantum circuit by implementing it as a subroutine.

Naively, one may expect that these protocols function in a modular fashion--if an alien gives us an unknown device that implements $U$, we would still be able to such protocols by treating this device as a black box. Just as we call upon the built-in functions in Mathematica, we can make use of $U$ without knowing anything about its physical implementation or circuit decomposition. This is useful computationally, since our procedure for computing the trace of $U$ is independent of the physical details of $U$. Indeed, in many physical situations such as in Hamiltonian estimation~\cite{Aharonov}, $U$ could represent some unknown physical process; and be poorly understood.

Here, we prove that if it were generally possible to execute unknown quantum subroutines, then a fundamental physical principle would be violated. This precludes any protocol that invokes a black-box unitary $U$ conditioned on a quantum mechanical state. The physical implementation of any such protocol, DQC1 included, must necessarily depend on the physical implementation of $U$. This calls into question whether it is still possible for quantum processors to offer any speed-up in computing certain properties of a black-box unitary.

We answer this question in presenting a \emph{black-box DQC1} protocol that efficiently evaluates $|\tr(U)|$, even when $U$ represents an unknown physical process; such that only its input-output relations are accessible. Thus, we introduce a class of protocols that demonstrate exponential speed-up on black-box inputs. We demonstrate that this modularity has immediate practical consequence. A polynomial sequence of black-box DQC1 protocols imparts full quantum factoring capability with efficiency comparable to existing quantum factoring algorithms~\cite{Shor:1994jg, Kitaev1995, Plenio}. This enhanced factoring algorithm displays the many advantages of modularity, in both a quadratic reduction in the number of control gates required for implementation, and in reducing the extent to which quantum circuits need to be tailored to factor specific numbers.

\begin{figure*}[Htb!]
\begin{center}
\includegraphics[width=0.9\textwidth]{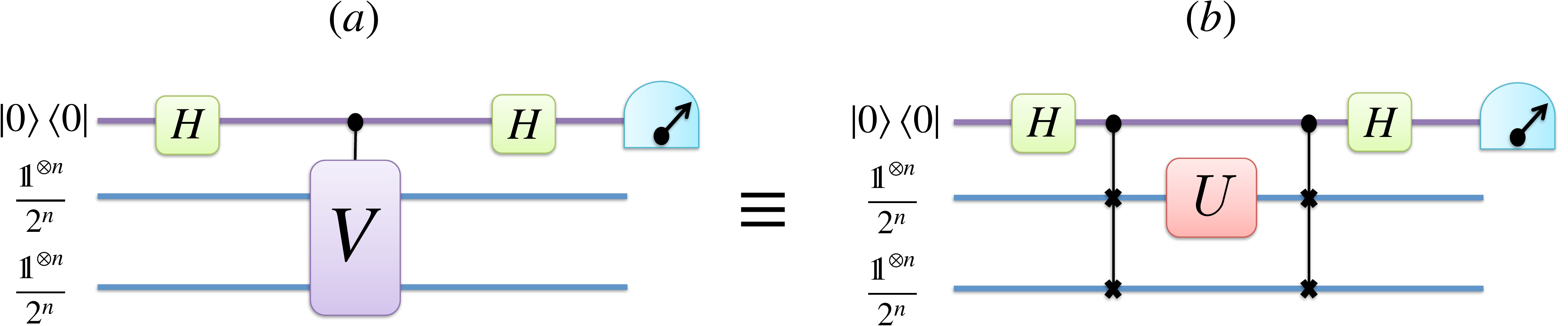}
\end{center}
 \caption{\textbf{DQC1 and black-box DQC1.}The standard DQC1 protocol (a) operates by implementing a unitary $V$ on a register of completely mixed qubits, controlled an an ancillary qubit in the state $H \ket{0} = (\ket{0}+\ket{1})/\sqrt{2}$ (The control unitary $V_c = \ket{0}\bra{0}\otimes \openone+ \ket{1}\bra{1}\otimes V$ takes a state $\ket{x}\ket{\phi}$ to $\ket{x}V^x\ket{\phi}$, where $x = \{0,1\}$). Appropriate measurements of the ancillary qubit allows evaluation of $\tr(V)$. The black-box DQC1 circuit (b) using a unitary $U$ implements a special case of of the standard DQC1 protocol when $V = U \otimes U^{\dagger}$. This allows evaluation of $\tr(V) = |\tr(U)|^2$.
 \label{fig:dqcp}}
\end{figure*}

\section{Results}

\textbf{Executing black-box subroutines.} The evaluation of $|\tr(U)|$ can, of course, be systematically solved by DQC1 when one is given a physical decomposition of $U$ in terms of elementary quantum gates. In this paradigm, one can implement $U$ on a register of $n$ completely mixed qubits, controlled on the state of a single pure qubit (See Fig.~\ref{fig:dqcp}.a). Polynomial repeated measurements of this qubit in the $X$ and $Y$ basis can respectively give an estimate of the real and complex components of $\tr(U)$ to any arbitrary fixed accuracy ~\cite{Knill, DQC1}. Clearly, taking the modulus of these estimates allows accurate estimation of $|\tr(U)|$.

\begin{figure}
\begin{center}
\includegraphics[width=0.45\textwidth]{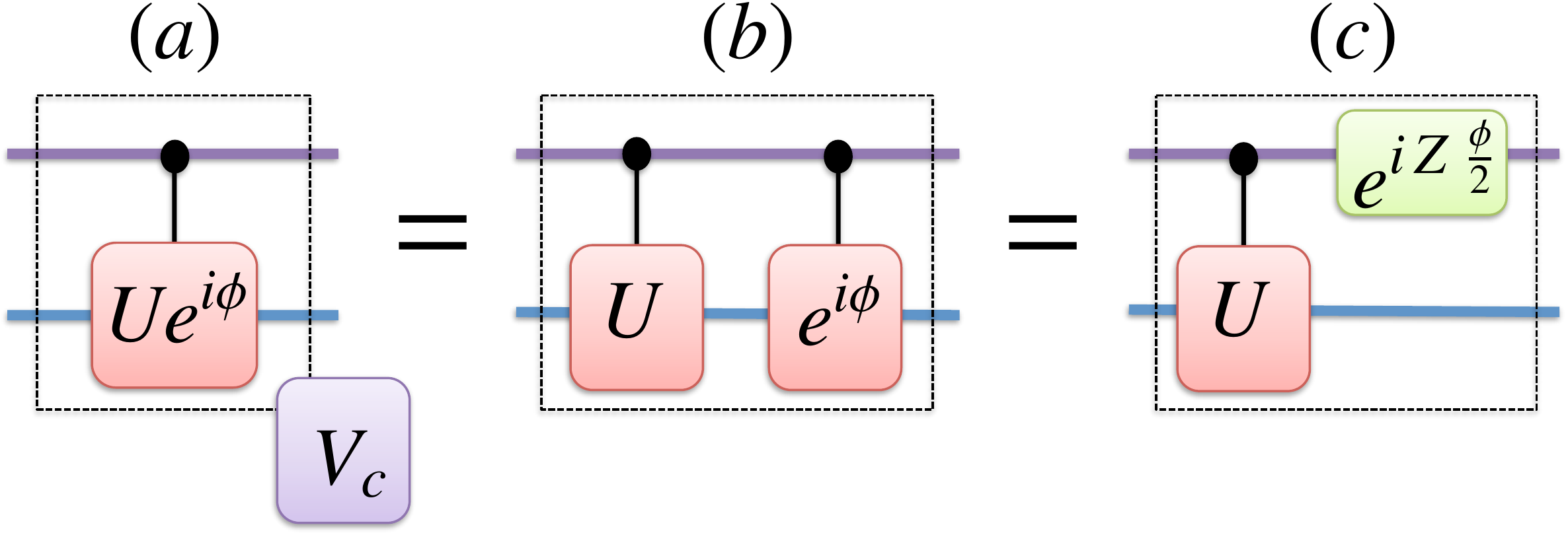}
\end{center}
\caption{\textbf{Non-physicality of the Global Phase.}
One of the standard methods to observe the global phase of a unitary $V$ is to implement $V_c$, i.e., control $V$ with respect to some external degree of freedom (a). Suppose $V = e^{i\phi}U$ for some fixed $U$. The control of $V$ can be decomposed into a control on $U$, followed by a control of $e^{i\phi}$ (b). However, the control of $e^{i\phi}$ is actually mathematically equivalent to a local operation on the control qubit (c). We see that the controlled global phase is a physical property of the control, rather than the system being controlled. It is therefore impossible to extract the global phase information from a black-box unitary.\label{fig:controlunitary}}
\end{figure}

The general DQC1 protocol, however, no longer functions in situations where $U$ is a \emph{black-box unitary}, and represents the actions of a completely unknown physical process. In this scenario, no amount of repeated experiments should ever reveal to us any information about the global phase of $U$. Indeed, if any protocols existed that could differentiate two black box unitary transformations, $U$ and $e^{i \phi}U$, that differ only by a global phase, it would violate one of the fundamental tenets of quantum theory: two unitary operations that differ only by a global phase represent the same physical process. Thus, \emph{any quantum algorithm whose output depends on the global phase $\phi$ cannot be implemented when supplied with a black-box unitary}. Indeed, locality considerations indicate that the global phase is not stored within an unknown physical process, and thus cannot be measured (See Fig.~\ref{fig:controlunitary}). Note that this argument also applies when many copies of $U$ are supplied.

Noting that the trace of $U$ and $e^{i \phi}U$ differ, we conclude that \emph{computing the trace of a black-box unitary is impossible}. To see exactly where the standard DQC1 protocol fails, we note that DQC1 relies on adding a control to the unitary operator $U$. This procedure would automatically reveal information about the global phase, and is thus not implementable when $U$ represents a completely unknown physical process.

Indeed, existing methods for adding controls to unknown physical processes depend crucially on extra knowledge that these processes act as the identity on some subspace of the input system ~\cite{Zhou, zhou_unknown}. Such methods creatively exploit special properties in certain physical systems, such as linear optics, and thus require knowledge of the physical processes that underly $U$ (See Methods for an extended discussion). When such knowledge is unavailable, this is no longer possible; and we will need to abandon adding controls to unknown unitary transformations to restore modularity.

\begin{figure}
\includegraphics[width=0.3\textwidth]{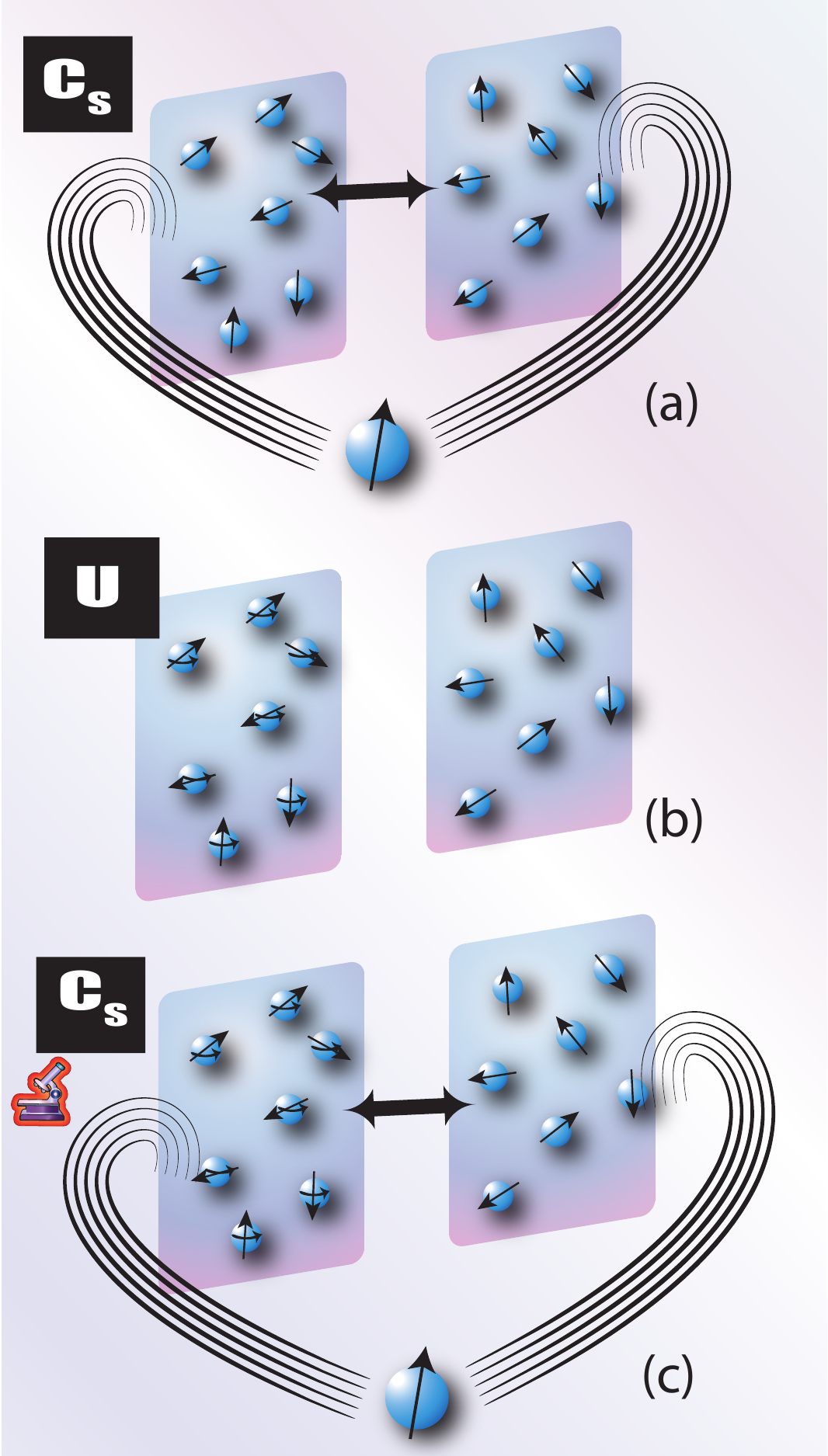}
\caption{\label{fig:operationalinterpretation} \textbf{The black-box DQC1 circuit} can be operationally interpreted by considering two boxes full of completely mixed qubits which are (a) swapped controlled on a single external pure qubit (This is formally a controlled-\textsc{swap} gate, $S_c$, where $S \ket{\phi}\ket{\psi} = \ket{\psi}\ket{\phi}$ is the \textsc{swap} gate that swaps the states of the two boxes, see supplementary materials for details). The black-box unitary $U$ is then performed on the qubits in one of the boxes (b); this operation leaves the box completely mixed. A second controlled-\textsc{swap} is then performed followed by a measurement on the external control qubit (c).}
\end{figure}

\textbf{Black-Box DQC1.} The modulus of the trace of a unitary contains no information about its global phase, and thus evaluation of $|\tr(U)|$ is not forbidden by the above constraints. We propose a \emph{black-box DQC1 protocol} that performs this task in Fig.~\ref{fig:operationalinterpretation}. In this protocol, we begin with a pure control qubit and \emph{two} completely mixed registers of $n$ qubits. The two registers are then coherently swapped, by controlling on the state of an ancillary qubit. While the usefulness of this operation appears highly paradoxical--swapping two completely mixed subsystems that look identical appears to achieve little--it is in fact the only interaction between control and register that we need. $|\mathrm{Tr}[U]|$ can be evaluated by feeding one of the registers into the black box that implements $U$, together with controlled-\textsc{swap} operations. This protocol involves only a single pure qubit, and therefore falls within the DQC1 paradigm. The protocol is operationally equivalent to executing standard DQC1 to compute the trace of $V = U^{\dag} \otimes U$ (See Fig.~\ref{fig:dqcp}.a). Measurement of the control qubit thus estimates $\tr(V) = |\tr(U)|^2$.

The black-box DQC1 protocol makes no sacrifices on efficiency; while it requires a doubling in the size of the maximally mixed register, this is counterbalanced by the need to only measure the control qubit in a single basis. Meanwhile, it features a notable advantage; in dropping the requirement for us to have any knowledge of $U$, black-box DQC1 functions as a algorithm that truly treats $U$ as an arbitrary input. The quantum circuit for the protocol does not need to be individually tailored to specific inputs; we can design an optimal implementation on the controlled-\textsc{swap} gates, and use the resulting design to probe the modulus of the trace of an arbitrary physical process.

The above results indicate that when we do not care about the global phase of $U$, modularity can sometimes be restored. One may speculate whether this sacrifice makes the protocol trivial; perhaps classical algorithms of comparable efficiency exist. Certain appeals to intuition appear compelling. The \textsc{swap} operation is Hermitian, and control Hermitian operators have been suspected to be classical within the DQC1 setting. They generate no quantum correlations, entanglement or discord~\cite{Dakic, DQC1, Vedral, Zurek}. Meanwhile, $U$ is applied locally, without modification, on a maximally mixed register; and does not facilitate interactions between control and register at all. We will see however, that the same protocol can be applied to construct a modular variant of Shor's algorithm.

\textbf{Modular Factoring with black-box DQC1.} A polynomial sequence of black-box DQC1 circuits can factor efficiently. The core information required to factor can be encoded in the eigenspectrum of a specific Hamiltonian, $H_a$. In general this energy eigenspectrum is of the form $ \{j/r : a^r \equiv 1 \, {\rm mod} \, N, j = 0,\dots, r-1\}$, while the ability to find $r$ such that $a^r \equiv\,1 \,{\rm mod} \, N$ for some suitable $a < N$ is sufficient to find a factor of $N = pq$.

These eigenvalues can be isolated by implementing a series of unitary operations, $U_a^{2^x} = \exp{( 2^x 2 \pi i H_a)}$, for \, $x =0,1, \dots, L -1 \in O(\log_2 N)$ on a completely mixed register of $n = \log_2 N$ qubits, controlled on some ancillary qubit; followed by an inverse quantum Fourier transform (analogous to standard phase estimation protocols). The value of $r$ can then be retrieved with high probability~\cite{Plenio}.

\begin{figure*}
\begin{center}
\includegraphics[width=0.9\textwidth]{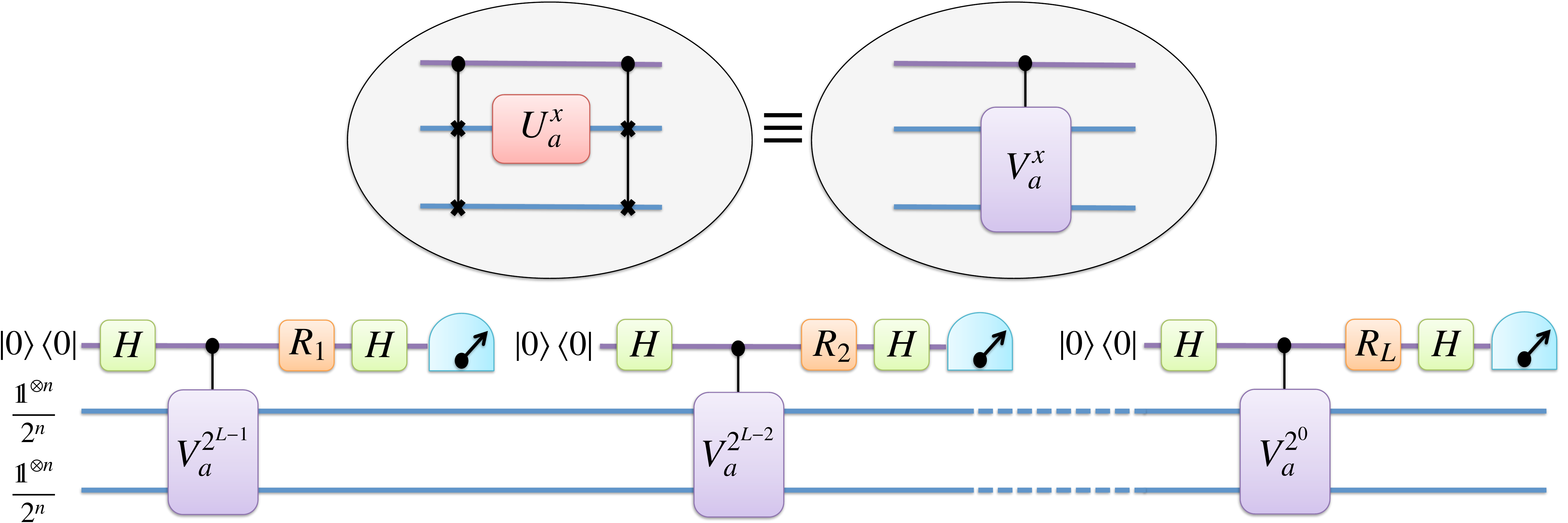}
\end{center}
 \caption{We can factor with a polynomial number of copies of the black-box DQC1 protocol; explicitly for the inputs to the factoring circuit each copy of the black-box DQC1 subroutine will be function equivalently to a controlled-$V^x = U^x \otimes U^{x\dag}$. The remaining Hadamard and $R_j$ gates are tantamount to a quantum Fourier transform modulo $ 2^L$ on the control qubit's state, where $L \in O(\log_2 N)$. The operator $R_j = |0\rangle\langle 0| +|1\rangle \langle 1| \exp{(- 2\pi i\sum_k m_{j-k}/2^k)} $ is applied to the control where $m_{j-k} = (1,0)$ such that the value $1$ is assigned if the $(j-k)^{th}$ detector clicked and $0$ if it did not; the index $k$ runs over the previous detectors.\label{fig:factoringsinglequbit}}
\end{figure*}

The black-box DQC1 architecture can be used instead of a control unitary operator, as the elementary building block of a factoring routine (See Fig.~\ref{fig:factoringsinglequbit}). The resulting algorithm recovers the differences between eigenvalues; in general these are also of the form $k/r$ for $k = 0, \dots, r-1$; for the purposes of factoring these differences contain the same amount of useful information as the spectrum itself. In Methods and Supplementary Materials, we prove our construction succeeds in approximately
\begin{gather}\label{eq:efficency}
O\left(\frac{pq }{(p-1)(q-1)} \log{\log{r}}\right)
\end{gather}
runs. This is comparable to Shor's algorithm which typically succeeds in $O(\log{\log{r}})$ runs.

This results in a much more modular implementation of Shor's protocol. Conventional methods would require decomposing each unitary $U_a^{2^x}$ into a sequence of elementary quantum gates, and explicitly implementing individual controls on each gate. This would require $O(n^3)$ controls per operator \cite{Vlatko}, and force a redesign of the circuit when factoring different numbers (as $U_a$ depends on $N$). In contrast, in the black-box enhanced factoring algorithm, the number of controls can be significantly reduced - we need only $O(n)$ controlled-\textsc{swap} gates, all of which can be reused regardless of which number we factor. This modularity may improve the feasibility of non-compiled quantum factoring protocols; making it possible to go beyond the current criticism that pre-existing compiled implementations use prior knowledge of the answer to simplify the computation~\cite{smolin}.

\section{Discussion}

In this article, we explored a general class of quantum protocols in which the input is not a quantum state, but rather a physical process that implements a unitary $U$. We demonstrated that when the only accessible information about the process is its black-box properties, any protocols that reveal the global phase of $U$ will fail. This constraint effects many existing quantum protocols, including quantum phase estimation and DQC1; these algorithms would at best, need to be tailored to each specific unitary, and at worst, cease to function. This immediately motivates the question, whether these protocols can be modified, such that modularity is restored.

We addressed this question in proposing a black-box DQC1 protocol, designed to function even when $U$ is supplied as an unknown physical process. In executing $U$ purely as a subroutine, the protocol is able to evaluate $|\tr(U)|$ of an exponentially sized $U$, in polynomial time. A polynomial sequence of these protocols allows efficient factoring. This establishes that certain quantum algorithms do not require exact execution of a desired sub-routine to achieve its intended output.

The results presented are of both practical and foundational interest. In the theory of computation, the modularity of algorithms is a feature of great importance. Most complex algorithms consists of many different subroutines, and it is highly desirable if these algorithms can be constructed separately, and combined by a third party who needs not understand the exact code, circuit, or physical implementation of each individual component. Our results thus provide a no-go theorem that details when such modularity is impossible, and simultaneously presents a method to avoids this situation for certain computational problems.

Our method can be used to construct black-box variants of other quantum protocols whose input is encoded within a controlled unitary operator; candidates include quantum phase estimation,  as well as quantum algorithms to solve linear systems, and simulate Jarzynski's equality~\cite{metrology, Kitaev1995,Loyd_Linear,linear_exp,dorner2013}. In some scenarios, as was in DQC1, this extra modularity comes at the price of sacrificing some of the information such algorithms compute. In other's however, such as factoring, the information sacrificed may not be of any relevance, and thus modularity comes for free.

From a foundational viewpoint, the study of quantum computation has identified a number of classically trivial tasks whose quantum analogues are impossible~\cite{wootters1982,dieks1982,kumar2011quantum}. The observation that certain properties of $U$ can never be computed from its black-box properties adds to this surprising list. This seems to mirror the Halting problem (and its generalization to Rice's theorem~\cite{rice}) in computer science, which indicates that not all non-trivial black-box properties of an algorithm can be computed from its code. Here, quantum theory hints at a converse; not all properties of an algorithm's code can be computed from its black-box properties.

\section{Methods}

\textbf{Control unitary transformations in special physical architectures.}
There exist situations where partial information is available about a unitary $U$, such as what physical processes underly its synthesis. Here, $U$ is not longer a black-box unitary, and this extra knowledge can help us construct algorithms that determine the global phase of $U$ and/or add a control to $U$.

Let $\mathcal{P}$ denote a physical process that synthesizes an unknown $U \in SU(d)$ that acts on input system $\mathcal{S}$. $\mathcal{S}$ thus encodes a quantum state $\ket{\phi}$ spanned by $\ket{0},\ldots, \ket{d-1}$, which transforms to $U\ket{\phi}$ under the action of $\mathcal{P}$. Our no-go result states that \emph{With no prior knowledge of $\mathcal{P}$, computing the global phase of $U$ is impossible}, and as a corollary, we cannot add a control to $U$.

However, if we (i) can isolate an extra $d$ degrees of freedom, $\ket{d},\ldots,\ket{2d-1}$ in $\mathcal{S}$, (ii) know that $\mathcal{P}$ leaves these degrees of freedom unchanged, i.e., $\mathcal{P}: \ket{d+k} \rightarrow \ket{d+k}$ for all $k = 0,\ldots,d-1$. The action of $\mathcal{P}$ on the entire $2d$ degrees of freedom in $S$ has the matrix representation
\begin{gather}
V = \left(\begin{array}{cc} U & 0 \\ 0 & \openone_d \\ \end{array} \right)
\end{gather}
where $\openone_d$ is an identity matrix of dimension $d$. We can define a `virtual' qubit by relabeling $\ket{d*b +k }$ as $\ket{k}_A \ket{b}_B$ where $b \in \{0,1\}$, then $V$ coincides exactly with applying a unitary $U$ to $A$ controlled on $B$. This strategy relied on knowledge of both (i) and (ii) and thus $U$ is not a black-box. Formally, the phase of $U$ is not a global phase, and our no-go result does not apply.

Note that \emph{ancillary qubits do not help}. If $\mathcal{P}$ acts only on a qudit system $\mathcal{A}$, and we introduced an ancillary qubit $\mathcal{B}$, then the action of $\mathcal{P}$ on this joint system is represented by the unitary $U \otimes \openone$, which does not equal $U \oplus \openone$.

However, certain physical architectures allow constructions that satisfy (i) and (ii). For example, let $\mathcal{S}$ be an optical mode and let $\mathcal{P}$ represent a sequence of linear optical gates that implements some $U \in SU(2)$ on the polarization degree of freedom of $\mathcal{S}$. Physics tells us that $U$ does not affect the vacuum. Thus, the action of $\mathcal{P}$ is in fact, $U$ controlled on whether a photon is present in the mode. This enables adding controls on an unknown $U$ using photons traveling in a superposition of two different optical paths, one of which passes though $\mathcal{P}$~\cite{Zhou,zhou_unknown}. Of course, such methods depend crucially on prior knowledge that $\mathcal{P}$ does not decohere the path degree of freedom, and fails if $U$ is truly a black box; for example, when $U$ is implemented by some third party, who chooses to encode output $U\ket{\phi}$ in a different physical system. Upon the completion of this work, we were notified of a complementary viewpoint of these experiments which was concurrently developed by Brukner et al~\cite{caslav}.

\textbf{Intuition behind Modular Factoring}. To efficiently factor, it is sufficient to have an efficient algorithm that solves the order finding problem~\cite{Shor:1994jg, Kitaev1995}: Given an input $a \in {\Bbb N}$, $1 < a < N$, output the first value of $r$ such that $a^r \equiv 1 \, \mod N$. If $a$ is chosen at random, the value of $r$ will, with good probability, reveal the factors of $N$.

Quantum factoring algorithms function by noting the eigenvalues of a modular exponentiation operator
 \begin{gather}\label{eq:modularexponentiation}
 U_a|x \rangle = |(x * a ) \,{\rm mod} \, N \rangle,
\end{gather}
encode the value of $r$. Since $\left( U_a\right)^r$ is the $N$-dimensional identity matrix. This last constraint forces the eigenvalues of this operator to be the $r^{th}$ roots of unity; these are complex numbers $\omega^{-j}$ which carry information about $r$ through $\omega = \exp{2 \pi i /r}$ and $j = 1,\dots, r-1$. If we can measure the phase of an eigenvalue for which $j/r$ is an irreducible fraction then we can find $r$.

Due to the closure of the $r^{th}$ roots of unitary under multiplication the eigenvalues of $U_a \otimes U_a^{\dagger}$ are also $r^{th}$ roots of unity. Hence it is functionally equivalent to find the phase associated with an eigenvalue of $V_a = U_a \otimes U_a^{\dagger}$. Thus, noting that the black-box DQC1 protocol with input $U_a$ is equivalent to DQC1 with input $V_a$, we may replace each control $U_a$ with its more modular variant with negligible loss in efficiency (See proof in Supplementary Materials).

\textbf{Acknowledgements.} The authors would like to thank A. Brodutch, O. Dalhston, and C. Weedbrook for discussions. This work is supported by the John Templeton Foundation, the National Research Foundation and Ministry of Education in Singapore, the Academic Research Fund Tier 3 MOE2012-T3-1-009, the National Basic Research Program of China Grant 2011CBA00300, 2011CBA00302 and the National Natural Science Foundation of China Grant 61033001, 61061130540. After the completion of our work, we were notified that the black-box DQC1 algorithm is also being explored in a different context~\cite{Mio}.

\section{Supplementary material}

\textbf{Proof of Correctness}. Firstly we characterize the modular exponentiation operator defined in Eq.~\eqref{eq:modularexponentiation}. Every eigenvector of this $N \otimes N $ unitary operator can be expressed in terms of some natural number $g_d < N$, as:

\begin{align}
\label{eq:eigenvector}
|\psi_{j_d}\rangle = \frac{1}{\sqrt{r_d}} \left({\omega_d}^{-j_d(1)} |g_d*a\rangle + \dots + {\omega_d}^{-j_d(r_d)}|g_d*a^{r_{d}}\rangle\right)\notag
\end{align}
where $r_d$ is an exponent satisfying $g_d*a^{r_d} \equiv g_d \mod N$ while the coefficients are defined through $\omega_d = \exp{ 2\pi i /r_d }$ and $j_d \in \{0,\dots, r_d -1 \}$. The associated eigenvalue is $\omega_d^{j_d}$. Note that the case $g_d = 1$ has $r$ eigenvectors and associated eigenvalues of the form $\omega^j = \exp{2\pi i j /r}$ for $j = 0,\dots, r-1$, while in general $r_d | r $ because $a^r \equiv 1 \, {\rm mod }\, N$. Furthermore whenever $N = pq$ is coprime with $g_d$ the relation $g_d(a^{r_d} - 1) \equiv 1 \, {\rm mod} \, N $ implies $r_d = r$; these conditions are met by $(p-1)(q-1)$ natural numbers less than $N$. Implying that at most $p +q -1$ possible values of $g_d$ correspond to eigenrelations for $U_a$ where the phase of $\omega_d$ has denominator $r_d \neq r$ \cite{Plenio}.

With respect to the eigenbasis $|\psi_{j_d}\rangle$ we write the operator $U_a$ as
\begin{gather}
U_a = \sum_d\sum_{[j_d=0,\dots, r_d-1]} w_d^{j_d} |\psi_{j_d}\rangle\langle \psi_{j_d}|,
\end{gather}
where the first sum, indexed by $d$, runs over the set $\{g_d\}$ and the nested sum runs over $ j_d = 0,\dots, r_d-1$.

We now use this information to analyse the circuit in Fig.~\ref{fig:factoringsinglequbit}. We simplify the calculation by using the binary decimal expansion
\begin{gather}
0.c_l c_{l+1}\dots c_m = \frac{1}{2} c_l + \frac{1}{4}c_{l+1} +\dots + \frac{1}{2^{m-l+1}} c_m.
\end{gather}

In this convention a measurement of the control register at the end of the circuit yields a number
\begin{gather}
c = \sum_{i=0}^{L-1} 2^i c_i
\end{gather}
where the binary digit $c_i$ is 1 if the {\it ith} detector clicked and $0$ otherwise, while the binary decimal $c/2^L$ is the best estimate to some eigenvalue of $U_a \otimes U_a^{\dagger}$. To achieve sufficient accuracy we require $L = \log_2 t$ ancillary qubits where $t$ is the power of $2$ satisfying $N^2 \le t \le 2 N^2$ \cite{Plenio}.

The probability of obtaining a specific binary number $c$ when measuring the circuit in Figure~\ref{fig:factoringsinglequbit} is:
\begin{gather}\label{eq:probabilityofchosinganelement}
P(c) = \frac{1}{N^2 t^2}\sum_{d, d'} \sum_{j_d, j_d'}|G|^2,
\end{gather}
 where
 \begin{gather}
 \qquad G = \sum_{b = 0}^{t-1} \exp\left(2\pi b i \left(\frac{j_d}{r_d} - \frac{j_d'}{r_d'} - \frac{c}{t} \right)\right).
\end{gather}
We deliberately chose the number of control qubits so that our measurement $c/t$ can resolve $j_d/r_d - j_d'/r_d'$ to an accuracy sufficient for determining $r$: this implies their exists an eigenvalue for which our estimate has a bounded amount of error:
\begin{gather}\label{eq:conditionsforbound}
\left|\frac{j_d}{r_d} - \frac{j_d'}{r_d'} - \frac{c}{t}\right| \le \frac{1}{2t}.
\end{gather}
Under these conditions we inherit a lower bound on $|G|^2 \ge 4 t^2/ \pi^2 $~\cite{Plenio}, see also~\cite{Shor:1994jg} for a more detailed argument.

If we are going to be successful in retrieving any information about $r$ from $c/t$ then (a) we need $j_d/r_d - j_d'/r_d'$ to have denominator $r$ and (b) we need the numerator to be coprime with $r$.

Firstly there are $(p-1)(q-1)$ values of $g$ which are coprime with $N$ permitting at least $(p-1)(q-1)/r$, values of $r_d = r$~\cite{Plenio}. For each value of $r_d = r$ the number of eigenvalues corresponding to irreducible fractions $j/r$ where $j \in \{0,\dots, r-1\}$ is defined through Euler's totient function $\phi(r)$; which follows the relation $\phi(r)/r > \delta/\log{\log{r}}$ for a constant $\delta$ \cite{Hardy, Shor:1994jg, Plenio}.

In the next section we demonstrate that for every $j_d/r_d$ satisfying $r_d = r$ and $gcd(j_d, r) = 1$ there is a faction $j_d/r_d - j_d'/r_d'$ satisfying both (a) and (b); by symmetry this argument should apply equally to $j_d'/r_d'$. Hence the number of eigenvalues $j_d/r_d - j_d'/r_d'$ from which we can successfully determine $r$ is:
\begin{gather}\label{eq:numberofsuccessfulleigenvalues}
Num_{c} = N^2 - \left( N - \chi \right)^2 = \chi (2N - \chi)
\end{gather}
where $\chi = \frac{\phi(r) (p-1)(q-1)}{r}$. And the probability our circuit succeeds (that is estimates a fraction with denominator $r$ and numerator coprime with $r$) is
\begin{gather}
P'(c) = Num_c * P(c) \ge \frac{4 t^2}{N^2 t^2 \pi^2}\, \chi \left(2N - \chi \right)
\end{gather}
For a direct comparison with Shor's result~\cite{Shor:1994jg} we give the lower bound on the success probability:
\begin{gather}
\frac{4}{N \pi^2} (p-1)(q-1)\frac{\phi(r)}{r}.
\end{gather}
This scales as the same order in $N$ as standard factoring algorithms~\cite{Shor:1994jg,Plenio}; in fact, asymptotically the probability of success using the black-box DQC1 protocol goes like $P * (2 - P)$ where $P$ is the probability of success for Parker and Plenio's factoring routine~\cite{Plenio}; so to first order in $P$ (which tends to $0$ as $N \rightarrow \infty$) we get a doubling in the success probability of the black-box DQC1 protocol over that of Parker and Plenio, which recovers the cost of the extra register qubits used in our construction.

\section{The number of fraction $j_d/r_d - j_d'/rd'$ which have denominator $r$ and coprime numerator}\label{sec:countingargument}

This section contains information required to derive Eq.~\eqref{eq:numberofsuccessfulleigenvalues}.

Firstly fix the eigenvalue $j_d'/r_d'$ and assume $r_d = r$ then
\begin{gather}\label{eq:denominatorequaltor}
j_d/r_d - j_d'/r_d' = \frac{j - k_d' j_d'}{r},
\end{gather}
where we have let $r_d' = r/k_d'$ for some integer $k_d'$ (which is always possible because $r_d'$ divides $r$).

Now for a fixed value of $j_d'/r_d'$ there are $r$ possible numerators in Eq.~\eqref{eq:denominatorequaltor} corresponding to the possible values of $j = 0,\dots, r-1$. We want to establish a one to one correspondence between values of $j$ which are coprime with $r$ and values of the numerator of Eq.~\eqref{eq:denominatorequaltor} which are coprime with $r$ (for a fixed $j_d'/r_d'$).

Since we have fixed $k_d' j_d'$ we know $j - k_d'j_d' \equiv 0,\dots, r-1 \mod r$ (i.e., when $j = 0, \dots, r -1$ so does $j - k_d'j_d' \, {\rm mod} \, r$).

Additionally for any $\alpha, \beta \in {\Bbb Z}$ we have: $\alpha +\beta*r$ is coprime with $r$ if and only if $\alpha$ is coprime with $r$ (this follows very quickly from the contrapositive).

We show the relation in one direction $\alpha +\beta*r \textrm{ coprime with } r \longrightarrow \alpha \textrm{ coprime with } r$ using the contrapositive. First assume $\alpha$ shares a common factor with $r$; that is let $\alpha = \lambda*\tau$ and $r = \lambda*\kappa$ (for integers $\kappa,\, \tau,\,\lambda$). This implies $\alpha + \beta*r = \lambda (\tau + \beta *\kappa)$ and therefore $\alpha +\beta*r$ is not coprime with $r$. $\square$

It follows that for a fixed $j_d'/r_d'$; if the conditions: (a) fraction has denominator $r$ and (b) numerator is coprime with $r$, are satisfied by $j_d/r_d$ then there is a corresponding value of $j_d/r_d - j_d'/r_d'$ also satisfying (a) and (b). This argument is symmetric and can also be applied to $j_d'/r_d'$.

So the number of eigenvalues $j_d/r_d - j_d'/r_d'$ which can not be used to determine $r$ is the number of pairs $(j_d/r_d, j_d'/r_d')$ for which is it impossible to determine $r$ from either $j_d/r_d$, or $j_d'/r_d'$. Eq.~\eqref{eq:numberofsuccessfulleigenvalues} is simply the total number of eigenvalues of $U_a \otimes U_a^{\dagger}$ minus the number that can not be used to determine $r$.

\section{Equivalence of the black-box DQC1 subroutine to a control unitary on a completely mixed register}

\begin{figure*}[tb]
\begin{center}
\includegraphics[width=0.9\textwidth]{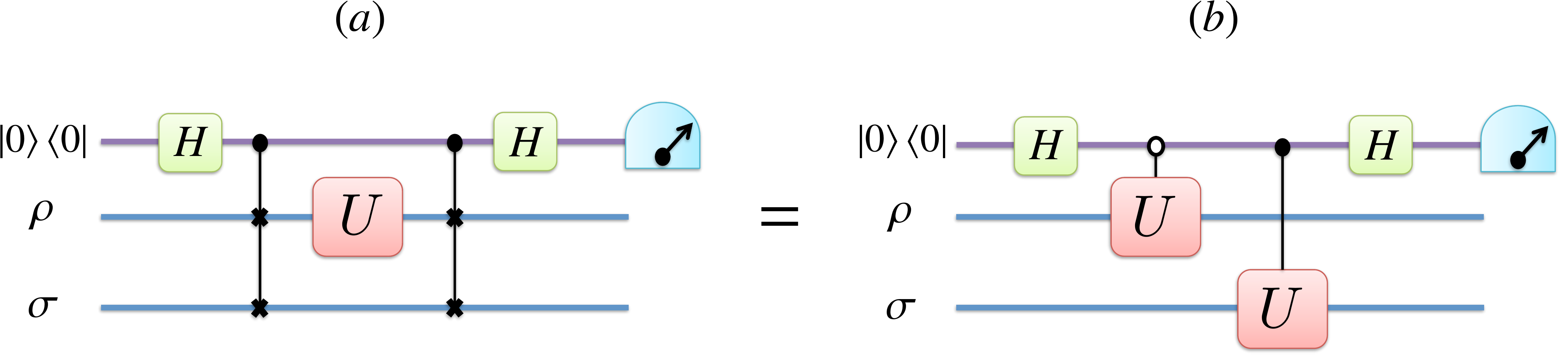}
\end{center}
 \caption{\textbf{Black-box DQC1 in a control unitary architecture.} A black-box DQC1 circuit (a) is formally equivalent to a pair of controlled unitary transformations: $|0\rangle\langle 0| \otimes U \otimes \openone+ |1\rangle \langle 1 | \otimes \openone \otimes \openone$ and $|0\rangle \langle 0| \otimes \openone \otimes \openone + |1\rangle \langle 1| \otimes \openone \otimes U$, respectively shown in (b). This equivalence holds for an arbitrary $\rho$, $\sigma$, and $U$. The measurement of the control qubit at the end yields the value of $\tr[U \rho] \times \tr[\sigma U^\dag]$. \label{fig:foramlblack-boxDQC1equivalence}}
\end{figure*}

We characterize the resulting action of the black-box DQC1 protocol for the general case where the two reservoirs are initialized in arbitrary states $\rho$ and $\sigma$. We compare this to the use of a control $U \otimes U^{\dagger}$ operation. We demonstrate equivalence when the case of the factoring protocol.

Consider two states $\ket{\psi}= \sum_{ijk} \alpha_{ijk} \ket{i,j,k}$ and $\ket{\phi}= \sum_{ijk} \beta_{ijk} \ket{i,j,k}$ which are each composed of three qubits.
The tensor product of these states is
\begin{gather}
\ket{\psi}\otimes \ket{\phi}= \sum_{ijk} \sum_{lmn} \alpha_{ijk} \beta_{lmn} \ket{il,jm,kn}.
\end{gather}
We define the operator $S$ which swaps the $m$th qubit of $\psi$ with $m$th qubit of $\phi$:
\begin{align}
S \ket{\psi}\otimes \ket{\phi} =& S \sum_{ijk} \sum_{lmn} \alpha_{ijk} \beta_{lmn} \ket{il,jm,kn}\\
=& \sum_{ijk} \sum_{lmn} \alpha_{ijk} \beta_{lmn} \ket{li,mj,nk}\\
=& \ket{\phi}\otimes \ket{\psi}.
\end{align}

This furnishes a \textsc{swap} operator which interchanges two $m$ qubit registers $\rho$ and $\sigma$
\begin{gather}
\rho \otimes \sigma \to S \rho \otimes \sigma S = \sigma \otimes \rho.
\end{gather}

When the registers are initialized as two arbitrary $m$ qubit states, $\rho$ and $\sigma$, due to the relation $U~\otimes~\openone_m~\rho~\otimes~\sigma~S~U^{\dagger}~\otimes~\openone_m~S~=~U~\rho~\otimes~\sigma~U^{\dagger}$, the state of the black-box DQC1 circuit after the second \textsc{swap} in Figure~\ref{fig:dqcp} is
\begin{align}\label{eqn:blasted}
\tau_{BB} = & \frac{1}{2^{2m+1}}
\begin{pmatrix} \rho \otimes \sigma & U \rho \otimes \sigma U^\dag \cr
\rho U^\dag \otimes U \sigma & \rho \otimes U\sigma U^{\dag} \cr\end{pmatrix}.
\end{align}

When $\rho$ and $\sigma$ are eigenstates of $U$ with eigenvalues $e^{i\lambda_{\rho}}$ and $e^{i\lambda_{\sigma}}$ respectively then the final state of the circuit is

\begin{align}
\tau_{BB} = & \frac{1}{2^{2m+1}}
\begin{pmatrix} \rho \otimes \sigma & e^{i(\lambda_{\rho} - \lambda_{\sigma})} \rho \otimes \sigma \cr
e^{i(\lambda_{\sigma} -\lambda_{\rho})} \rho \otimes \sigma & \rho \otimes \sigma \cr\end{pmatrix}.
\end{align}
By comparison the state of a circuit implementing a controlled-$U \otimes U^{\dagger}$ on two registers initialized as $\rho$ and $\sigma$ is:
\begin{align}
\tau_{U \otimes U^{\dag}} = & \frac{1}{2^{2m+1}}
\begin{pmatrix} \rho \otimes \sigma & \rho U \otimes \sigma U^{\dag} \cr
U^{\dag} \rho \otimes U \sigma & U^\dag \rho U\otimes U\sigma U^\dag \cr\end{pmatrix}.
\end{align}
In general the final state of these circuits are the same when the registers are initialized as eigenstates of $U$:
\begin{align}
\tau_{U \otimes U^{\dag}} = & \frac{1}{2^{2m+1}}
\begin{pmatrix} \rho \otimes \sigma & e^{i(\lambda_{\rho} - \lambda_{\sigma})} \rho \otimes \sigma \cr
e^{i(\lambda_{\sigma} -\lambda_{\rho})} \rho \otimes \sigma &\rho \otimes \sigma \cr\end{pmatrix}.
\end{align}
Due to the linearity of quantum mechanics, the two circuits are equal for any input state that is an improper mixture of eigenstates of $U$, i.e; any density operator that is diagonal in the eigenbasis of $U$. This clearly includes complete mixed states, and all inputs during the operation of the modular factoring algorithm.

In the most general case, the black-box DQC1 circuit as represented in Eq. (\ref{eqn:blasted}) is formally equivalent to a pair of controlled unitary transformations as outlined in Fig.~\ref{fig:foramlblack-boxDQC1equivalence}.


\begin{thebibliography}{28}
\expandafter\ifx\csname natexlab\endcsname\relax\def\natexlab#1{#1}\fi
\expandafter\ifx\csname bibnamefont\endcsname\relax
  \def\bibnamefont#1{#1}\fi
\expandafter\ifx\csname bibfnamefont\endcsname\relax
  \def\bibfnamefont#1{#1}\fi
\expandafter\ifx\csname citenamefont\endcsname\relax
  \def\citenamefont#1{#1}\fi
\expandafter\ifx\csname url\endcsname\relax
  \def\url#1{\texttt{#1}}\fi
\expandafter\ifx\csname urlprefix\endcsname\relax\def\urlprefix{URL }\fi
\providecommand{\bibinfo}[2]{#2}
\providecommand{\eprint}[2][]{\url{#2}}

\bibitem[{\citenamefont{Knill and Laflamme}(1998)}]{Knill}
\bibinfo{author}{\bibfnamefont{E.}~\bibnamefont{Knill}} \bibnamefont{and}
  \bibinfo{author}{\bibfnamefont{R.}~\bibnamefont{Laflamme}},
  \bibinfo{journal}{Physical Review Letters} \textbf{\bibinfo{volume}{81}},
  \bibinfo{pages}{5672} (\bibinfo{year}{1998}).

\bibitem[{\citenamefont{Harrow et~al.}(2009)\citenamefont{Harrow, Hassidim, and
  Lloyd}}]{Loyd_Linear}
\bibinfo{author}{\bibfnamefont{A.~W.} \bibnamefont{Harrow}},
  \bibinfo{author}{\bibfnamefont{A.}~\bibnamefont{Hassidim}}, \bibnamefont{and}
  \bibinfo{author}{\bibfnamefont{S.}~\bibnamefont{Lloyd}},
  \bibinfo{journal}{Physical Review Letters} \textbf{\bibinfo{volume}{103}},
  \bibinfo{pages}{150502} (\bibinfo{year}{2009}).

\bibitem[{\citenamefont{Datta et~al.}(2005)\citenamefont{Datta, Flammia, and
  Caves}}]{DQC1}
\bibinfo{author}{\bibfnamefont{A.}~\bibnamefont{Datta}},
  \bibinfo{author}{\bibfnamefont{S.~T.} \bibnamefont{Flammia}},
  \bibnamefont{and} \bibinfo{author}{\bibfnamefont{C.~M.} \bibnamefont{Caves}},
  \bibinfo{journal}{Physical Review A} \textbf{\bibinfo{volume}{72}},
  \bibinfo{pages}{042316} (\bibinfo{year}{2005}).

\bibitem[{\citenamefont{Ekert et~al.}(2002)\citenamefont{Ekert, Alves, Oi,
  Horodecki, Horodecki, and Kwek}}]{dqc1-1}
\bibinfo{author}{\bibfnamefont{A.~K.} \bibnamefont{Ekert}},
  \bibinfo{author}{\bibfnamefont{C.~M.} \bibnamefont{Alves}},
  \bibinfo{author}{\bibfnamefont{D.~K.~L.} \bibnamefont{Oi}},
  \bibinfo{author}{\bibfnamefont{M.}~\bibnamefont{Horodecki}},
  \bibinfo{author}{\bibfnamefont{P.}~\bibnamefont{Horodecki}},
  \bibnamefont{and} \bibinfo{author}{\bibfnamefont{L.~C.} \bibnamefont{Kwek}},
  \bibinfo{journal}{Phys. Rev. Lett.} \textbf{\bibinfo{volume}{88}},
  \bibinfo{pages}{217901} (\bibinfo{year}{2002}).

\bibitem[{\citenamefont{Poulin et~al.}(2003)\citenamefont{Poulin, Laflamme,
  Milburn, and Paz}}]{dqc1-2}
\bibinfo{author}{\bibfnamefont{D.}~\bibnamefont{Poulin}},
  \bibinfo{author}{\bibfnamefont{R.}~\bibnamefont{Laflamme}},
  \bibinfo{author}{\bibfnamefont{G.~J.} \bibnamefont{Milburn}},
  \bibnamefont{and} \bibinfo{author}{\bibfnamefont{J.~P.} \bibnamefont{Paz}},
  \bibinfo{journal}{Phys. Rev. A} \textbf{\bibinfo{volume}{68}},
  \bibinfo{pages}{022302} (\bibinfo{year}{2003}).

\bibitem[{\citenamefont{Emerson et~al.}(2004)\citenamefont{Emerson, Lloyd,
  Poulin, and Cory}}]{dqc1-3}
\bibinfo{author}{\bibfnamefont{J.}~\bibnamefont{Emerson}},
  \bibinfo{author}{\bibfnamefont{S.}~\bibnamefont{Lloyd}},
  \bibinfo{author}{\bibfnamefont{D.}~\bibnamefont{Poulin}}, \bibnamefont{and}
  \bibinfo{author}{\bibfnamefont{D.}~\bibnamefont{Cory}},
  \bibinfo{journal}{Phys. Rev. A} \textbf{\bibinfo{volume}{69}},
  \bibinfo{pages}{050305} (\bibinfo{year}{2004}).

\bibitem[{\citenamefont{Poulin et~al.}(2004)\citenamefont{Poulin, Blume-Kohout,
  Laflamme, and Ollivier}}]{dqc1-4}
\bibinfo{author}{\bibfnamefont{D.}~\bibnamefont{Poulin}},
  \bibinfo{author}{\bibfnamefont{R.}~\bibnamefont{Blume-Kohout}},
  \bibinfo{author}{\bibfnamefont{R.}~\bibnamefont{Laflamme}}, \bibnamefont{and}
  \bibinfo{author}{\bibfnamefont{H.}~\bibnamefont{Ollivier}},
  \bibinfo{journal}{Phys. Rev. Lett.} \textbf{\bibinfo{volume}{92}},
  \bibinfo{pages}{177906} (\bibinfo{year}{2004}).

\bibitem[{\citenamefont{Aharonov et~al.}(2002)\citenamefont{Aharonov, Massar,
  and Popescu}}]{Aharonov}
\bibinfo{author}{\bibfnamefont{Y.}~\bibnamefont{Aharonov}},
  \bibinfo{author}{\bibfnamefont{S.}~\bibnamefont{Massar}}, \bibnamefont{and}
  \bibinfo{author}{\bibfnamefont{S.}~\bibnamefont{Popescu}},
  \bibinfo{journal}{Physical Review A} \textbf{\bibinfo{volume}{66}},
  \bibinfo{pages}{052107} (\bibinfo{year}{2002}).

\bibitem[{\citenamefont{Shor}(1997)}]{Shor:1994jg}
\bibinfo{author}{\bibfnamefont{P.~W.} \bibnamefont{Shor}},
  \bibinfo{journal}{SIAM journal on computing} \textbf{\bibinfo{volume}{26}},
  \bibinfo{pages}{1484} (\bibinfo{year}{1997}).

\bibitem[{\citenamefont{Kitaev}(1995)}]{Kitaev1995}
\bibinfo{author}{\bibfnamefont{A.~Y.} \bibnamefont{Kitaev}},
  \bibinfo{journal}{arXiv:quant-ph/9511026}  (\bibinfo{year}{1995}).

\bibitem[{\citenamefont{Parker and Plenio}(2000)}]{Plenio}
\bibinfo{author}{\bibfnamefont{S.}~\bibnamefont{Parker}} \bibnamefont{and}
  \bibinfo{author}{\bibfnamefont{M.}~\bibnamefont{Plenio}},
  \bibinfo{journal}{Physical Review Letters} \textbf{\bibinfo{volume}{85}},
  \bibinfo{pages}{3049} (\bibinfo{year}{2000}).

\bibitem[{\citenamefont{Zhou et~al.}(2011)\citenamefont{Zhou, Ralph, Kalasuwan,
  Zhang, Peruzzo, Lanyon, and O'Brien}}]{Zhou}
\bibinfo{author}{\bibfnamefont{X.-Q.} \bibnamefont{Zhou}},
  \bibinfo{author}{\bibfnamefont{T.~C.} \bibnamefont{Ralph}},
  \bibinfo{author}{\bibfnamefont{P.}~\bibnamefont{Kalasuwan}},
  \bibinfo{author}{\bibfnamefont{M.}~\bibnamefont{Zhang}},
  \bibinfo{author}{\bibfnamefont{A.}~\bibnamefont{Peruzzo}},
  \bibinfo{author}{\bibfnamefont{B.~P.} \bibnamefont{Lanyon}},
  \bibnamefont{and} \bibinfo{author}{\bibfnamefont{J.~L.}
  \bibnamefont{O'Brien}}, \bibinfo{journal}{Nature communications}
  \textbf{\bibinfo{volume}{2}}, \bibinfo{pages}{413} (\bibinfo{year}{2011}).

\bibitem[{\citenamefont{Zhou et~al.}(2013)\citenamefont{Zhou, Kalasuwan, Ralph,
  and O'Brien}}]{zhou_unknown}
\bibinfo{author}{\bibfnamefont{X.-Q.} \bibnamefont{Zhou}},
  \bibinfo{author}{\bibfnamefont{P.}~\bibnamefont{Kalasuwan}},
  \bibinfo{author}{\bibfnamefont{T.~C.} \bibnamefont{Ralph}}, \bibnamefont{and}
  \bibinfo{author}{\bibfnamefont{J.~L.} \bibnamefont{O'Brien}},
  \bibinfo{journal}{Nature Photonics} pp. \bibinfo{pages}{223---228}
  (\bibinfo{year}{2013}).

\bibitem[{\citenamefont{Daki{\'c} et~al.}(2010)\citenamefont{Daki{\'c}, Vedral,
  and Brukner}}]{Dakic}
\bibinfo{author}{\bibfnamefont{B.}~\bibnamefont{Daki{\'c}}},
  \bibinfo{author}{\bibfnamefont{V.}~\bibnamefont{Vedral}}, \bibnamefont{and}
  \bibinfo{author}{\bibfnamefont{{\v{C}}.}~\bibnamefont{Brukner}},
  \bibinfo{journal}{Physical Review Letters} \textbf{\bibinfo{volume}{105}},
  \bibinfo{pages}{190502} (\bibinfo{year}{2010}).

\bibitem[{\citenamefont{Henderson and Vedral}(2001)}]{Vedral}
\bibinfo{author}{\bibfnamefont{L.}~\bibnamefont{Henderson}} \bibnamefont{and}
  \bibinfo{author}{\bibfnamefont{V.}~\bibnamefont{Vedral}},
  \bibinfo{journal}{Journal of Physics A: mathematical and general}
  \textbf{\bibinfo{volume}{34}}, \bibinfo{pages}{6899} (\bibinfo{year}{2001}).

\bibitem[{\citenamefont{Ollivier and Zurek}(2001)}]{Zurek}
\bibinfo{author}{\bibfnamefont{H.}~\bibnamefont{Ollivier}} \bibnamefont{and}
  \bibinfo{author}{\bibfnamefont{W.~H.} \bibnamefont{Zurek}},
  \bibinfo{journal}{Physical Review Letters} \textbf{\bibinfo{volume}{88}},
  \bibinfo{pages}{017901} (\bibinfo{year}{2001}).

\bibitem[{\citenamefont{Vedral et~al.}(1996)\citenamefont{Vedral, Barenco, and
  Ekert}}]{Vlatko}
\bibinfo{author}{\bibfnamefont{V.}~\bibnamefont{Vedral}},
  \bibinfo{author}{\bibfnamefont{A.}~\bibnamefont{Barenco}}, \bibnamefont{and}
  \bibinfo{author}{\bibfnamefont{A.}~\bibnamefont{Ekert}},
  \bibinfo{journal}{Physical Review A} \textbf{\bibinfo{volume}{54}},
  \bibinfo{pages}{147} (\bibinfo{year}{1996}).

\bibitem[{\citenamefont{Smolin et~al.}(2013)\citenamefont{Smolin, Smith, and
  Vargo}}]{smolin}
\bibinfo{author}{\bibfnamefont{J.~A.} \bibnamefont{Smolin}},
  \bibinfo{author}{\bibfnamefont{G.}~\bibnamefont{Smith}}, \bibnamefont{and}
  \bibinfo{author}{\bibfnamefont{A.}~\bibnamefont{Vargo}},
  \bibinfo{journal}{Nature} \textbf{\bibinfo{volume}{499}},
  \bibinfo{pages}{163} (\bibinfo{year}{2013}).

\bibitem[{\citenamefont{Giovannetti et~al.}(2006)\citenamefont{Giovannetti,
  Lloyd, and Maccone}}]{metrology}
\bibinfo{author}{\bibfnamefont{V.}~\bibnamefont{Giovannetti}},
  \bibinfo{author}{\bibfnamefont{S.}~\bibnamefont{Lloyd}}, \bibnamefont{and}
  \bibinfo{author}{\bibfnamefont{L.}~\bibnamefont{Maccone}},
  \bibinfo{journal}{Physical Review Letters} \textbf{\bibinfo{volume}{96}},
  \bibinfo{pages}{010401} (\bibinfo{year}{2006}).

\bibitem[{\citenamefont{Cai et~al.}(2013)\citenamefont{Cai, Weedbrook, Su,
  Chen, Gu, Zhu, Li, Liu, Lu, and Pan}}]{linear_exp}
\bibinfo{author}{\bibfnamefont{X.-D.} \bibnamefont{Cai}},
  \bibinfo{author}{\bibfnamefont{C.}~\bibnamefont{Weedbrook}},
  \bibinfo{author}{\bibfnamefont{Z.-E.} \bibnamefont{Su}},
  \bibinfo{author}{\bibfnamefont{M.-C.} \bibnamefont{Chen}},
  \bibinfo{author}{\bibfnamefont{M.}~\bibnamefont{Gu}},
  \bibinfo{author}{\bibfnamefont{M.-J.} \bibnamefont{Zhu}},
  \bibinfo{author}{\bibfnamefont{L.}~\bibnamefont{Li}},
  \bibinfo{author}{\bibfnamefont{N.-L.} \bibnamefont{Liu}},
  \bibinfo{author}{\bibfnamefont{C.-Y.} \bibnamefont{Lu}}, \bibnamefont{and}
  \bibinfo{author}{\bibfnamefont{J.-W.} \bibnamefont{Pan}},
  \bibinfo{journal}{Phys. Rev. Lett.} \textbf{\bibinfo{volume}{110}},
  \bibinfo{pages}{230501} (\bibinfo{year}{2013}).

\bibitem[{\citenamefont{Dorner et~al.}(2013)\citenamefont{Dorner, Clark,
  Heaney, Fazio, Goold, and Vedral}}]{dorner2013}
\bibinfo{author}{\bibfnamefont{R.}~\bibnamefont{Dorner}},
  \bibinfo{author}{\bibfnamefont{S.}~\bibnamefont{Clark}},
  \bibinfo{author}{\bibfnamefont{L.}~\bibnamefont{Heaney}},
  \bibinfo{author}{\bibfnamefont{R.}~\bibnamefont{Fazio}},
  \bibinfo{author}{\bibfnamefont{J.}~\bibnamefont{Goold}}, \bibnamefont{and}
  \bibinfo{author}{\bibfnamefont{V.}~\bibnamefont{Vedral}},
  \bibinfo{journal}{Physical Review Letters} \textbf{\bibinfo{volume}{110}},
  \bibinfo{pages}{230601} (\bibinfo{year}{2013}).

\bibitem[{\citenamefont{Wootters and Zurek}(1982)}]{wootters1982}
\bibinfo{author}{\bibfnamefont{W.~K.} \bibnamefont{Wootters}} \bibnamefont{and}
  \bibinfo{author}{\bibfnamefont{W.~H.} \bibnamefont{Zurek}},
  \bibinfo{journal}{Nature} \textbf{\bibinfo{volume}{299}},
  \bibinfo{pages}{802} (\bibinfo{year}{1982}).

\bibitem[{\citenamefont{Dieks}(1982)}]{dieks1982}
\bibinfo{author}{\bibfnamefont{D.}~\bibnamefont{Dieks}},
  \bibinfo{journal}{Physics Letters A} \textbf{\bibinfo{volume}{92}},
  \bibinfo{pages}{271} (\bibinfo{year}{1982}).

\bibitem[{\citenamefont{Kumar and Paraoanu}(2011)}]{kumar2011quantum}
\bibinfo{author}{\bibfnamefont{K.~S.} \bibnamefont{Kumar}} \bibnamefont{and}
  \bibinfo{author}{\bibfnamefont{G.}~\bibnamefont{Paraoanu}},
  \bibinfo{journal}{EPL} \textbf{\bibinfo{volume}{93}}, \bibinfo{pages}{20005}
  (\bibinfo{year}{2011}).

\bibitem[{\citenamefont{Rice}(1953)}]{rice}
\bibinfo{author}{\bibfnamefont{H.~G.} \bibnamefont{Rice}},
  \bibinfo{journal}{Transactions of the American Mathematical Society}
  \textbf{\bibinfo{volume}{74}}, \bibinfo{pages}{358} (\bibinfo{year}{1953}).

\bibitem[{\citenamefont{Nakayama et~al.}(2013)\citenamefont{Nakayama, Soeda,
  and Murao}}]{Mio}
\bibinfo{author}{\bibfnamefont{S.}~\bibnamefont{Nakayama}},
  \bibinfo{author}{\bibfnamefont{A.}~\bibnamefont{Soeda}}, \bibnamefont{and}
  \bibinfo{author}{\bibfnamefont{M.}~\bibnamefont{Murao}},
  \bibinfo{journal}{arXiv:1310.3047}  (\bibinfo{year}{2013}).

\bibitem[{\citenamefont{Ara{\'u}jo et~al.}(2013)\citenamefont{Ara{\'u}jo, Feix,
  Costa, and Brukner}}]{caslav}
\bibinfo{author}{\bibfnamefont{M.}~\bibnamefont{Ara{\'u}jo}},
  \bibinfo{author}{\bibfnamefont{A.}~\bibnamefont{Feix}},
  \bibinfo{author}{\bibfnamefont{F.}~\bibnamefont{Costa}}, \bibnamefont{and}
  \bibinfo{author}{\bibfnamefont{{\v{C}}.}~\bibnamefont{Brukner}},
  \bibinfo{journal}{arXiv:1309.7976}  (\bibinfo{year}{2013}).

\bibitem[{\citenamefont{Hardy and Wright}(1979)}]{Hardy}
\bibinfo{author}{\bibfnamefont{G.~G.~H.} \bibnamefont{Hardy}} \bibnamefont{and}
  \bibinfo{author}{\bibfnamefont{E.~M.} \bibnamefont{Wright}},
  \emph{\bibinfo{title}{An introduction to the theory of numbers}}
  (\bibinfo{publisher}{Oxford University Press}, \bibinfo{year}{1979}).

\end{thebibliography}
\end{document}